\documentclass[prd,superscriptaddress,amsfonts,amssymb,amsmath,showpacs,onecolumn]{revtex4-2}
\usepackage{bm}
\usepackage{amsfonts}
\usepackage{latexsym}
\usepackage{graphicx}
\usepackage{amsmath}
\usepackage{palatino}
\usepackage{mathpazo}
\usepackage{textcomp}
\usepackage{float}
\usepackage{booktabs}
\usepackage{dcolumn}
\usepackage{booktabs}
\usepackage{multirow}
\usepackage{hyperref}
\hypersetup{colorlinks,citecolor=blue}
\usepackage{amsmath}
\usepackage{xcolor}
\usepackage{orcidlink}
\usepackage{epsfig}
\usepackage{caption}
\usepackage{subcaption}
\usepackage{commath}
\def\jnl@style{\it}
\def\aaref@jnl#1{{\jnl@style#1}}

\def\aaref@jnl#1{{\jnl@style#1}}

\def\aj{\aaref@jnl{AJ}}                   
\def\apj{\aaref@jnl{ApJ}}                 
\def\apjl{\aaref@jnl{ApJ}}                
\def\apjs{\aaref@jnl{ApJS}}               
\def\apss{\aaref@jnl{Ap\&SS}}             
\def\aap{\aaref@jnl{A\&A}}                
\def\aapr{\aaref@jnl{A\&A~Rev.}}          
\def\aaps{\aaref@jnl{A\&AS}}              
\def\mnras{\aaref@jnl{Mon.~Not.~Roy.~Astron.~Soc.}}             
\def\prd{\aaref@jnl{Phys.~Rev.~D}}        
\def\prc{\aaref@jnl{Phys.~Rev.~C}}  
\def\prl{\aaref@jnl{Phys.~Rev.~Lett.}}    
\def\qjras{\aaref@jnl{QJRAS}}             
\def\skytel{\aaref@jnl{S\&T}}             
\def\ssr{\aaref@jnl{Space~Sci.~Rev.}}     
\def\zap{\aaref@jnl{ZAp}}                 
\def\nat{\aaref@jnl{Nature}}              
\def\aplett{\aaref@jnl{Astrophys.~Lett.}} 
\def\apspr{\aaref@jnl{Astrophys.~Space~Phys.~Res.}} 
\def\physrep{\aaref@jnl{Phys.~Rep.}}      
\def\physscr{\aaref@jnl{Phys.~Scr}}       
\def\commat{\aaref@jnl{Comm.~Math.~Phys.}}              
\def\science{\aaref@jnl{Science}}               
\def\cqg{\aaref@jnl{Classical Quant.~Grav.}}            
\def\jpcs{\aaref@jnl{JPCS}}                                     
\def\ijmpd{\aaref@jnl{Int.~J.~Mod.~Phys.~D}}                    
\def\grg{\aaref@jnl{Gen.~Relat.~Gravit.}}               
\def\rpp{\aaref@jnl{Rep.~Prog.~Phys.}}          
\def\npa{\aaref@jnl{Nucl.~Phys.~A}}        
\def\lrr{\aaref@jnl{Living Rev.~Rel.}}                   
\def\jcap{\aaref@jnl{J.~Cosmology Astropart.~Phys.}}    
\def\rmp{\aaref@jnl{Rev.~Mod.~Phys.}}   
\def\epjc{\aaref@jnl{Eur.~Phys.~J.~C}}

\allowdisplaybreaks[1]

\begin{document}

\color{black}       

\title{Modified cosmology in $f(Q, L_m)$ gravity}

\author{Y. Myrzakulov\orcidlink{0000-0003-0160-0422}}\email[Email: ]{ymyrzakulov@gmail.com} 
\affiliation{Department of General \& Theoretical Physics, L.N. Gumilyov Eurasian National University, Astana, 010008, Kazakhstan.}

\author{Alnadhief H. A. Alfedeel\orcidlink{0000-0002-8036-268X}}%
\email[Email: ]{aaalnadhief@imamu.edu.sa}
\affiliation{Department of Mathematics and Statistics, Imam Mohammad Ibn Saud Islamic University (IMSIU),
Riyadh 13318, Saudi Arabia.}

\author{M. Koussour\orcidlink{0000-0002-4188-0572}}
\email[Email: ]{pr.mouhssine@gmail.com}
\affiliation{Department of Physics, University of Hassan II Casablanca, Morocco.}

\author{S. Muminov\orcidlink{0000-0003-2471-4836}}
\email[Email: ]{sokhibjan.muminov@gmail.com}
\affiliation{Mamun University, Bolkhovuz Street 2, Khiva 220900, Uzbekistan.}

\author{E. I. Hassan\orcidlink{0000-0000-0000-0000}}%
\email[Email: ]{eiabdalla@imamu.edu.sa}
\affiliation{Department of Mathematics and Statistics, Imam Mohammad Ibn Saud Islamic University (IMSIU),
Riyadh 13318, Saudi Arabia.}

\author{J. Rayimbaev\orcidlink{0000-0001-9293-1838}}
\email[Email: ]{javlon@astrin.uz}
\affiliation{New Uzbekistan University, Movarounnahr Street 1, Tashkent 100007, Uzbekistan.}

\begin{abstract}

In this letter, we investigate cosmology within the framework of modified $f(Q, L_m)$ gravity using the non-linear model $f(Q, L_m) = -Q + \alpha L_m^n + \beta$, where $\alpha$, $\beta$, and $n$ are free parameters. The modified Friedmann equations are derived for a matter-dominated universe, and an analytical solution is obtained. Using Hubble, Pantheon+, and joint datasets, we constrain the model parameters and examine their implications. The results indicate that the model accommodates varying $H_0$ values, contributing to the Hubble tension, while $n \neq 1$ suggests deviations from general relativity (GR). The deceleration parameter confirms a transition to acceleration at $z_t \approx 0.6 - 0.8$, with present values supporting cosmic acceleration. This model offers a viable alternative to GR-based cosmology.

\end{abstract}

\maketitle


\section{Introduction}\label{sec1}

Observational data from Type Ia supernovae (SNe Ia) \cite{Riess,Perlmutter}, combined with data from large-scale structures \cite{T.Koivisto,S.F.}, baryon acoustic oscillations (BAO) \cite{D.J.,W.J.}, and the cosmic microwave background radiation (CMBR) \cite{C.L.,R.R.}, strongly imply that the expansion of the universe is accelerating. In the standard cosmological model, this acceleration is attributed to dark energy (DE), with the cosmological constant $\Lambda$ being its simplest candidate \cite{S.W.}. However, despite its success in fitting observational data, the cosmological constant suffers from significant theoretical challenges, including the fine-tuning and coincidence problems \cite{E.J.}. This has motivated alternative approaches, such as modifications to Einstein’s general relativity (GR) on large scales. 

A widely studied class of modified gravity theories extends the Einstein-Hilbert action by introducing generic functions of geometric and matter-related quantities. Among these, $f(R)$ gravity \cite{Buchdahl/1970,Dunsby/2010,Carroll/2004} modifies GR by generalizing the Ricci scalar term, while other extensions include $f(R, L_m)$ gravity \cite{Harko/2010,Wang/2012,Goncalves/2023,Myrzakulova/2024,Myrzakulov/2024}, which incorporates matter interactions, and $f(G)$ gravity \cite{Felice/2009, Bamba/2017, Goheer/2009}, which involves the Gauss-Bonnet term. In addition, theories such as $f(R, \mathcal{T})$ gravity \cite{Harko/2011, Koussour_1/2022, Koussour_2/2022, Myrzakulov/2023, KK1} and $f(Q)$ gravity \cite{Jimenez/2018,Jimenez/2020,Khyllep/2021,MK1, MK2, MK3, MK4, MK5, MK6, MK7, MK8} have been proposed as alternatives that modify the gravitational interaction through non-metricity rather than curvature. A further generalization, $f(Q, \mathcal{T})$ gravity \cite{Xu/2019,Xu/2020,K6,K7,Bourakadi,KK2,KK3}, combines non-metricity with the trace of the energy-momentum tensor, allowing for richer phenomenological implications. In this study, we investigate the cosmology in $f(Q, L_m)$ gravity (where $Q$ is the non-metricity and $L_m$ is the matter Lagrangian), a modification that extends the non-metricity-based $f(Q)$ gravity by incorporating matter interactions. Several studies have explored the implications of $f(Q, L_m)$ gravity in cosmology, focusing on both theoretical developments and observational constraints. Hazarika et al. \cite{Hazarika/2024} provided a comprehensive analysis of the general framework of $f(Q, L_m)$ gravity, deriving its field equations and exploring its cosmological applications. Myrzakulov et al. \cite{fQL1} conducted an observational analysis of the late-time acceleration within this modified gravity using cosmological datasets to constrain model parameters. In a related study, Myrzakulov et al. \cite{fQL2} obtained analytical solutions for late-time cosmic evolution in $f(Q, L_m)$ gravity and compared them with observational data to assess their viability. Furthermore, Myrzakulov et al. \cite{fQL3} investigated the effects of bulk viscosity in this framework, demonstrating how dissipative processes influence cosmic expansion. Samaddar and Singh \cite{fQL4} introduced a novel approach to baryogenesis within $f(Q, L_m)$ gravity, highlighting its potential implications for early-universe physics. Collectively, these studies illustrate the versatility of $f(Q, L_m)$ gravity in addressing key cosmological questions, motivating further investigations into its theoretical foundations and observational signatures. 

The structure of this letter is as follows: In Sec. \ref{sec2}, we outline the fundamental framework of $f(Q, L_m)$ gravity, establishing the key equations governing the theory. In Sec. \ref{sec3}, we derive the modified Friedmann equations for a flat FLRW universe within this gravitational framework. Sec. \ref{sec4} introduces a specific cosmological model of $f(Q, L_m)$ gravity, where we obtain expressions for the Hubble parameter and the deceleration parameter. In Sec. \ref{sec5}, we constrain the model parameters using observational datasets, including Hubble, Pantheon+, and their combined analysis. In addition, we examine the evolution of the deceleration parameter based on the best-fit values derived from these datasets. Finally, in Sec. \ref{sec6}, we summarize our findings and discuss their implications for late-time cosmic evolution.

\section{$f(Q,L_{m})$ Gravity Theory}\label{sec2}

In this study, we investigate an extension of symmetric teleparallel gravity, where the gravitational action takes the form,
\begin{equation}
    S=\int f(Q,L_m) \sqrt{-g} d^4x, \label{Action}
\end{equation}
where $\sqrt{-g}$ is the determinant of the metric tensor, ensuring covariance under coordinate transformations. The function $f(Q, L_m)$ is an arbitrary function of the non-metricity scalar $Q$ and the matter Lagrangian $L_m$, introducing modifications beyond standard teleparallel and GR formulations.  

The non-metricity scalar $Q$ is defined as \cite{Jimenez/2018}
\begin{equation}
Q\equiv -g^{\mu \nu }(L_{\,\,\,\alpha \mu }^{\beta }L_{\,\,\,\nu \beta
}^{\alpha }-L_{\,\,\,\alpha \beta }^{\beta }L_{\,\,\,\mu \nu }^{\alpha }),
\label{2}
\end{equation}
where $L_{\alpha \gamma }^{\beta }$ is the disformation tensor, given explicitly by 
\begin{equation}
L_{\alpha \gamma }^{\beta }=\frac{1}{2}g^{\beta \eta }\left( Q_{\gamma
\alpha \eta }+Q_{\alpha \eta \gamma }-Q_{\eta \alpha \gamma }\right).  \label{3}
\end{equation}

The non-metricity tensor, which quantifies the failure of the metric to be covariantly conserved, is given by 
\begin{equation}
Q_{\gamma \mu \nu }=-\nabla _{\gamma }g_{\mu \nu }=-\partial _{\gamma
}g_{\mu \nu }+g_{\nu \sigma }\widetilde{\Gamma }{^{\sigma }}_{\mu \gamma
}+g_{\sigma \mu }\widetilde{\Gamma }{^{\sigma }}_{\nu \gamma },  \label{4}
\end{equation}%
where $\widetilde{\Gamma }{^{\gamma }}_{\mu \nu }$ denotes the connection in symmetric teleparallel geometry.  

The traces of the non-metricity tensor are defined as 
\begin{equation}
Q_{\beta }=g^{\mu \nu }Q_{\beta \mu \nu },\qquad \widetilde{Q}_{\beta
}=g^{\mu \nu }Q_{\mu \beta \nu }.  \label{5}
\end{equation}%

To facilitate the field equations, we introduce the superpotential tensor, also known as the non-metricity conjugate,  
\begin{equation}
\hspace{-0.5cm} P_{\ \ \mu \nu }^{\beta }\equiv \frac{1}{4}\bigg[-Q_{\ \
\mu \nu }^{\beta }+2Q_{\left( \mu \ \ \ \nu \right) }^{\ \ \ \beta
}+Q^{\beta }g_{\mu \nu }-\widetilde{Q}^{\beta }g_{\mu \nu }  
\hspace{-0.5cm} -\delta _{\ \ (\mu }^{\beta }Q_{\nu )}\bigg]=-\frac{1}{2}%
L_{\ \ \mu \nu }^{\beta }+\frac{1}{4}\left( Q^{\beta }-\widetilde{Q}^{\beta
}\right) g_{\mu \nu }-\frac{1}{4}\delta _{\ \ (\mu }^{\beta }Q_{\nu )}.\quad
\quad   \label{6}
\end{equation}

This tensor plays a role analogous to the contortion tensor in teleparallel gravity. Then, the non-metricity scalar is written in terms of this conjugate tensor as \cite{Jimenez/2018}
\begin{equation}
Q=-Q_{\beta \mu \nu }P^{\beta \mu \nu }=-\frac{1}{4}\big(-Q^{\beta \nu
\rho }Q_{\beta \nu \rho }+2Q^{\beta \nu \rho }Q_{\rho \beta \nu } 
-2Q^{\rho }\tilde{Q}_{\rho }+Q^{\rho }Q_{\rho }\big).  \label{7}
\end{equation}

By varying the action (\ref{Action}) with respect to the metric, we obtain the gravitational field equations,
\begin{equation}
\frac{2}{\sqrt{-g}}\nabla_\alpha(f_Q\sqrt{-g}P^\alpha_{\;\;\mu\nu}) +f_Q(P_{\mu\alpha\beta}Q_\nu^{\;\;\alpha\beta}-2Q^{\alpha\beta}_{\;\;\;\mu}P_{\alpha\beta\nu})
+\frac{1}{2}f g_{\mu\nu}=\frac{1}{2}f_{L_m}(g_{\mu\nu}L_m-T_{\mu\nu}),\label{field}
\end{equation}
where $f_Q = \partial f(Q, L_m) / \partial Q$ and $f_{L_m} = \partial f(Q, L_m) / \partial L_m$. 

For the special case $f(Q, L_m) = f(Q) + 2 L_m$, the equations reduce to those of $f(Q)$ gravity \cite{Jimenez/2018}. The energy-momentum tensor of the matter is given by  
\begin{equation}
    T_{\mu\nu}=-\frac{2}{\sqrt{-g}}\frac{\delta(\sqrt{-g}L_m)}{\delta g^{\mu\nu}}=g_{\mu\nu}L_m-2\frac{\partial L_m}{\partial g^{\mu\nu}},
\end{equation}

Again, varying the action with respect to the connection gives an additional equation,  
\begin{equation}
    \nabla_\mu\nabla_\nu\Bigl( 4\sqrt{-g}\,f_Q\,P^{\mu\nu}_{\;\;\;\;\alpha}+H_\alpha^{\;\;\mu\nu}\Bigl)=0,
\end{equation}
where $H_\alpha^{\;\;\mu\nu}$ represents the hypermomentum density,  
\begin{equation}
    H_\alpha^{\;\;\mu\nu}=\sqrt{-g}f_{L_m}\frac{\delta L_m}{\delta Y^\alpha_{\;\;\mu\nu}}.
\end{equation}

A key feature of $f(Q, L_m)$ gravity is the non-conservation of the energy-momentum tensor. Applying the covariant derivative to the field equations (\ref{field}) leads to  
\begin{equation}
D_\mu\,T^\mu_{\;\;\nu}= \frac{1}{f_{L_m}}\Bigl[ \frac{2}{\sqrt{-g}}\nabla_\alpha\nabla_\mu H_\nu^{\;\;\alpha\mu} + \nabla_\mu\,A^{\mu}_{\;\;\nu} - \nabla_\mu \bigr( \frac{1}{\sqrt{-g}}\nabla_\alpha H_\nu^{\;\;\alpha\mu}\bigr) \Bigr]=B_\nu \neq 0.
\end{equation}

This equation indicates that energy-momentum exchange occurs due to the interaction between geometry and matter. The non-conservation term $B_\nu$ depends on the function $f(Q, L_m) $, the form of $L_m$, and other dynamical variables.  

\section{Modified Friedmann Equations in $f(Q, L_m)$ Gravity}\label{sec3}

To explore the cosmology of $f(Q, L_m)$ gravity, we assume a homogeneous and isotropic universe described by the flat Friedmann-Lema\^{i}tre-Robertson-Walker (FLRW) metric \cite{ryden/2003}:
\begin{equation}
\label{FLRW}
    ds^2=-dt^2+a^2(t)(dx^2+dy^2+dz^2),
\end{equation}
where $a(t)$ is the cosmic scale factor. The Hubble parameter, $H = \dot{a}/a$, characterizes the universe's expansion rate. From this metric (\ref{FLRW}), the non-metricity scalar is given by $Q = 6H^2$.

The matter content is modeled as a perfect fluid with the energy-momentum tensor given by
\begin{equation}
    T_{\mu\nu}=(\rho+p)u_{\mu}u_{\nu}+pg_{\mu\nu},
    \label{EMT}
\end{equation}
where $\rho$ is the energy density, $p$ is the isotropic pressure, and $u_{\mu}$ represents the four-velocity.

Using Eqs. (\ref{FLRW}) and (\ref{EMT}), the gravitational field equations derived from the variation of the action yield the modified Friedmann equations \cite{Hazarika/2024,fQL1,fQL2,fQL3}:
\begin{eqnarray}
\label{F1}
    && 3H^2 =\frac{1}{4f_Q}\bigr[ f - f_{L_m}(\rho + L_m) \bigl],\\
   && \dot{H} + 3H^2 + \frac{\dot{f_Q}}{f_Q}H=\frac{1}{4f_Q}\bigr[ f + f_{L_m}(p - L_m) \bigl]. \label{F2}
\end{eqnarray}

The energy balance equation in $f(Q, L_m)$ gravity is given by
\begin{equation}
\dot{\rho} + 3 H (\rho + p) = B_\mu u^\mu,    
\label{bal}
\end{equation}
where the nonzero source term $B_\mu u^\mu$ implies that energy-momentum conservation does not hold in its conventional form. When $B_\mu u^\mu = 0$, standard conservation is recovered; otherwise, it accounts for an interaction between geometry and matter. Rearranging Eq. \eqref{F2} using Eq. \eqref{F1}, we obtain
\begin{equation}
2\dot{H}+3H^2=\frac{1}{4f_Q}\left[f+f_{L_m}\left(\rho+2p-L_m\right)\right]-2\frac{\dot{f}_Q}{f_Q}H.
\end{equation}

To facilitate Further analysis, the modified Friedmann equations can be rewritten in terms of an effective energy density and pressure,
\begin{equation}\label{41}
3H^2=\rho_{eff}, \ 2\dot{H}+3H^2=-p_{eff},
\end{equation}
where
\begin{eqnarray}
    && \rho_{eff} = \frac{1}{4f_Q}\big[ f - f_{L_m}(\rho + L_m) \big],
    \label{rho_eff}\\
    && p_{eff} = 2\frac{\dot{f_Q}}{f_Q}H - \frac{1}{4f_Q} \big[ f + f_{L_m}(\rho + 2p - L_m) \big].
    \label{p_eff}
\end{eqnarray}

Finally, the effective energy-momentum conservation equation in $f(Q, L_m)$ gravity is given by
\begin{equation}\label{cons1}
\dot{\rho}_{eff}+3H\left(\rho_{eff}+p_{eff}\right)=0.
\end{equation}

The implications of these equations will be explored in the subsequent sections, particularly in the context of observational constraints and cosmic evolution.

\section{Cosmological model}
\label{sec4}

In this study, we examine an extended functional form of $f(Q, L_m)$ gravity given by  
\begin{equation}
f(Q,L_m)=-Q + \alpha L_m^{n} + \beta,     
\end{equation}
where $\alpha$, $\beta$, and $n$ are free parameters. This model introduces a non-minimal coupling between the non-metricity scalar $Q$ and the matter Lagrangian $L_m$ through a power-law dependence on $L_m$. The choice of a non-linear matter coupling is motivated by the desire to go beyond the standard additive forms such as $f(Q) + 2 L_m$, and to explore richer cosmological dynamics. The parameter $n$ governs the deviation from the minimal coupling case: when $n = 1$, the model reduces to a linear coupling, while $n \ne 1$ introduces effective interactions between geometry and matter that can influence the cosmic expansion history. Such couplings have been widely studied in other modified gravity theories (e.g., $f(R, L_m)$ \cite{Harko/2010,Wang/2012,Goncalves/2023}, $f(R, L_m, T)$ \cite{Haghani/2021,Arora/2024,Loewer/2024}) and are known to generate significant phenomenological consequences, including possible deviations from standard matter conservation, modified effective equations of state, and late-time acceleration \cite{Myrzakulova/2024,Myrzakulov/2024}. Therefore, for this specific form of $f(Q, L_m)$ gravity, we assume the matter Lagrangian to be $L_m = \rho$ \cite{Harko/2015}. The modified Friedmann equations (\ref{F1}) and (\ref{F2}) for a matter-dominated universe take the form
\begin{eqnarray}
&& 3 H^2 = \frac{\alpha (2 n -1)}{2} \rho^n - \frac{\beta}{2},  
\label{FF1}\\
&& 3 H^2 + 2 \dot{H}= \frac{\alpha (n -1)}{2} \rho^n - \frac{\beta}{2}.   \label{FF2}
\end{eqnarray}

Specifically, when $\alpha = 2$, $\beta = 0$, and $n = 1$, the standard Friedmann equations of GR are recovered. From Eqs. (\ref{FF1}) and (\ref{FF2}), we obtain
\begin{equation}
\frac{dH}{dt} = \frac{3n}{2(1 - 2n)} H^2 + \frac{n\beta}{4(1 - 2n)}.    
\end{equation}

We can now rewrite the equation in terms of the redshift $z = \frac{a_0}{a(t)} - 1$, with the present-day scale factor given by $a_0 = 1$. Since $\dot{H} = \frac{dH}{dt}$, we use the relation $\frac{dt}{dz} = -\frac{1}{(1+z) H}$ to express $\dot{H} = -H (1+z) \frac{dH}{dz}$. This transforms the equation into,
\begin{equation}
 -H (1+z) \frac{dH}{dz}=\frac{3n}{2(1 - 2n)} H^2 + \frac{n\beta}{4(1 - 2n)}.    
\end{equation}

By integrating the above equation, the Hubble parameter can be expressed as a function of redshift, yielding the following expression:
\begin{equation}
H(z)= \left[ H_0^{2} (1+z)^{\frac{3n}{2n-1}} + \frac{\beta}{6} \left\{ (1+z)^{\frac{3n}{2n-1}} - 1 \right\} \right]^{\frac{1}{2}}.  
\label{Hz}
\end{equation}
where $H_0=H(z=0)$ denotes the current value of the Hubble parameter.

To understand the accelerated or decelerated behavior of the universe, we define the deceleration parameter $q$ as  
\begin{equation}
q = \frac{d}{dt} \left( \frac{1}{H} \right) - 1 = -\frac{\dot{H}}{H^2} - 1.    
\end{equation}

The sign of the deceleration parameter $q$ determines the nature of cosmic expansion. A positive $q$ ($q > 0$) corresponds to a decelerating universe, as seen in matter- and radiation-dominated eras. Conversely, a negative $q$ ($ q < 0$) indicates an accelerating expansion, characteristic of the current dark energy-dominated epoch. Using Eq. (\ref{Hz}), we obtain
\begin{equation}
q(z)=\frac{3 n \left(6 H_0^2+\beta\right)}{2 (2 n-1) \left[ 6 H_0^2+\beta\left\{1-  (1+z)^{\frac{3 n}{1-2 n}}\right\}\right]}-1.    
\end{equation}

In Eq. (\ref{Hz}), the Hubble parameter $H(z)$ is expressed as a function of redshift $z$, the model parameter $n$, the present Hubble value $H_0$, and the parameter $\beta$. Notably, the expression does not explicitly depend on $\alpha$, indicating that the evolution of $H(z)$ within this framework is independent of this parameter. This suggests that $\alpha$ does not play a direct role in determining the cosmic expansion history, as encoded in $H(z)$, but may instead influence other dynamical aspects of the model.

\section{Observational Data}
\label{sec5}

In this section, we present a statistical analysis using the Monte Carlo Markov Chain (MCMC) technique to assess the model's viability under consideration \cite{R26,R27}. The core aim is to compare theoretical predictions with available cosmological observations to determine the consistency of the model. Specifically, we focus on aligning the model’s predictions with observational Hubble data $H(z)$ and SNe Ia (Pantheon+) as key observational tests.

\subsection{Observational Hubble data}

For the Hubble parameter $H(z)$, we employ measurements derived from the cosmic chronometer (CC) method. This technique uses the differential age of galaxies as a means of estimating $H(z)$, and the dataset includes 31 carefully compiled data points \cite{Moresco/2015}. The statistical comparison is carried out using the $\chi^2$ function:
\begin{equation}
\chi^2_{Hubble}= \sum_{i=1}^{31} \frac{\left[H(z_{i}) - H_{obs}(z_{i})\right]^2}{\sigma(z_{i})^{2}},
\end{equation}
where $\sigma(z_{i})$ represents the error in the observational data at redshift $z_{i}$, and $H_{obs}(z_{i})$ is the observed value of the Hubble parameter at this redshift.

\subsection{Pantheon+ data}

The Pantheon+ dataset, a more recent compilation of SNIa, contains 1701 individual measurements of apparent magnitudes over a broad range of redshifts, from $z = 0.001$ to $z = 2.3$. This extensive dataset presents vital details about the evolution of cosmic expansion, significantly advancing our understanding beyond previous datasets such as Union \cite{R30}, Union \cite{R31, R32}, JLA \cite{R33}, Pantheon \cite{R34}, and Pantheon+ \cite{R35}. SNe Ia are crucial in cosmology due to their well-understood luminosity, allowing them to be treated as standard candles for measuring cosmic distances. The statistical treatment of the Pantheon+ data is performed using the $\chi^2$ function:
\begin{equation}\label{4c}
\chi^2_{SNe}=  D^T C^{-1}_{SNe} D,
\end{equation}
where $C_{SNe}$ is the covariance matrix that accounts for both statistical and systematic uncertainties in the data. The vector $D$ is defined as
\begin{equation}
D = m_{Bi} - M - \mu^{th}(z_i),
\end{equation}
where $m_{Bi}$ is the apparent magnitude, $M$ represents the absolute magnitude, and $\mu^{th}(z_i)$ is the theoretical distance modulus predicted by the model, which can be expressed as
\begin{equation}\label{4d}
\mu^{th}(z_i) = 5 \log_{10} \left( \frac{D_L(z_i)}{1 \text{Mpc}} \right) + 25.
\end{equation}

Here, $D_L(z)$ is the luminosity distance, which is computed within the context of the MOG model using the integral:
\begin{equation}\label{4e}
D_L(z) = c(1 + z) \int_{0}^{z} \frac{dz'}{H(z', \theta)},
\end{equation}
where $\theta$ represents the model parameters, and $H(x, \theta)$ is the Hubble parameter at redshift $z'$. To address the degeneracy between the Hubble constant $H_0$ and the absolute magnitude $M$, the Pantheon+ dataset adopts a refined method for the vector $D$, defined as
\begin{equation}\label{4f}
\bar{D} = \begin{cases}
     m_{Bi} - M - \mu_i^{Ceph} & i \in \text{Cepheid hosts}, \\
     m_{Bi} - M - \mu^{th}(z_i) & \text{otherwise}.
    \end{cases}   
\end{equation}

In this approach, $\mu_i^{Ceph}$ is independently estimated using Cepheid variables as calibrators, enhancing the accuracy of the distance modulus. Thus, the $\chi^2$ function becomes $\chi^2_{SNe} = \bar{D}^T C^{-1}_{SNe} \bar{D}$.

The $\chi^2$ function for the combined $H(z)$ and Pantheon+ datasets is expressed as $\chi^2_{total}=\chi^2_{Hubble}+\chi^2_{SNe}$. The corner plot of the parameters is shown in Fig. \ref{F_Com}, while the numerical results are summarized in Tab. \ref{tab}.

\subsection{Results}

\begin{figure}[H]
\centering
\includegraphics[scale=0.57]{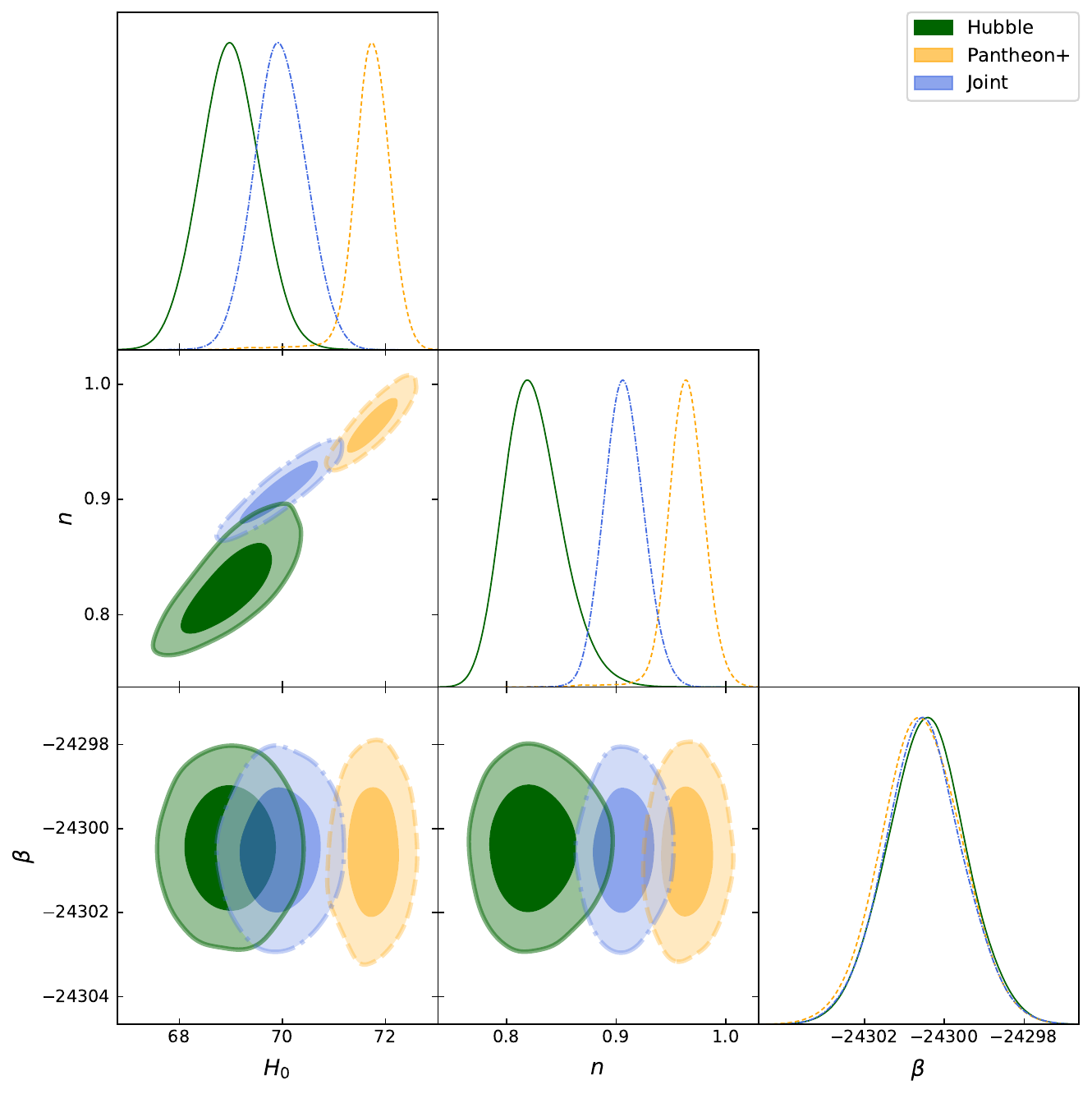}
\caption{The figure represents the corner plot for the parameter constraints in the $f(Q,L_m)$ model using Hubble, Pantheon+, and joint datasets.}
\label{F_Com}
\end{figure}

Fig. \ref{F_Com} presents the corner plots for the $f(Q,L_m)$ model in a matter-dominated universe, showing posterior distributions and confidence contours for $H_0$, $n$, and $\beta$ using different datasets. The Hubble dataset gives $H_0 = 68.99 \pm 0.58$ and $n = 0.825^{+0.022}_{-0.030}$, indicating a deviation from GR ($n=1$), while the Pantheon+ dataset yields a higher $H_0 = 71.73^{+0.35}_{-0.31}$ and $n = 0.964 \pm 0.018$, making it more consistent with GR. The joint analysis finds $H_0 = 69.96 \pm 0.50$ and $n = 0.907 \pm 0.018$, balancing both datasets. The parameter $\beta$ remains stable across all cases ($\sim -24300$), suggesting robustness. The model accommodates different $H_0$ values, potentially addressing the Hubble tension, while the deviation of $n$ from 1 implies a modification to GR-based matter evolution, particularly in the Hubble dataset. The correlations indicate a moderate link between $H_0$ and $n$, while $\beta$ remains largely independent. When comparing these values of $H_0$ with literature, our results are slightly higher than the Planck (2020) value of $H_0 = 67.4 \pm 0.5 \, \text{km/s/Mpc}$ \cite{Planck/2020}, particularly the Pantheon+ dataset. The Pantheon+ value is also somewhat closer to the SH0ES (2021) measurement of $H_0 = 73.2 \pm 1.3 \, \text{km/s/Mpc}$ \cite{Riess/2022}, but still within the lower range of its uncertainty. On the other hand, our joint dataset value of $H_0 = 69.96 \, \text{km/s/Mpc}$ aligns very closely with the recent work by Cao and Ratra \cite{Cao/2023}, which gives $H_0 = 69.8 \pm 1.3 \, \text{km/s/Mpc}$, indicating a good match with other late-universe measurements. Generally, our results contribute to the ongoing Hubble tension debate, as they fall between early-universe (Planck) and late-universe (SH0ES) measurements, highlighting the persistent discrepancy between these two approaches.

\begin{table}[h]
    \centering
    \begin{tabular}{|c|c|c|c|}
        \hline
        Dataset & $H_0$ (km/s/Mpc) & $n$ & $\beta$ \\  
        \hline
        Hubble & $68.99 \pm 0.58$ & $0.825^{+0.022}_{-0.030}$ & $-24300.45 \pm 0.99$ \\  
        \hline
        Pantheon+ & $71.73^{+0.35}_{-0.31}$ & $0.964 \pm 0.018$ & $-24300.6 \pm 1.0$ \\  
        \hline
        Joint & $69.96 \pm 0.50$ & $0.907 \pm 0.018$ & $-24300.51 \pm 0.98$ \\  
        \hline
    \end{tabular}
    \caption{The table represents the best-fit values of the parameters for the model $f(Q,L_m)$ using Hubble, Pantheon+, and joint datasets.}
    \label{tab}
\end{table}

\begin{figure}[h]
\centering
\includegraphics[scale=0.8]{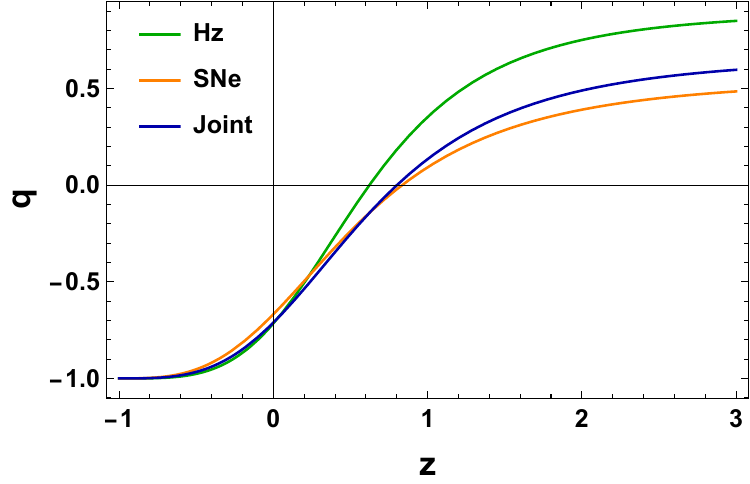}
\caption{The figure represents the evolution of the deceleration parameter $q$ as a function of redshift $z$ for the $f(Q,L_m)$ model, constrained using Hubble, Pantheon+, and joint datasets.}
\label{F_q}
\end{figure}

Fig. \ref{F_q} presents the evolution of the deceleration parameter $q$ as a function of redshift $z$ for the $f(Q,L_m)$ model, with constraints from Hubble (green), Pantheon+ (orange), and joint datasets (blue). At high redshift ($z > 2$), all curves converge to positive values, indicating a decelerated expansion consistent with a matter-dominated era. The transition from deceleration to acceleration occurs around $z_t \approx 0.6,0.8$, varying slightly across datasets. At $z = 0$, the present deceleration parameter remains negative, confirming cosmic acceleration. For $z = -1$, all curves tend toward $q \approx -1$, suggesting a de Sitter-like future evolution. The joint dataset provides an intermediate behavior between Hubble and Pantheon+, ensuring a balanced transition history. The present values of $q_0$ are $q_0 \approx -0.716$ for the Hubble dataset, $q_0 \approx -0.668$ for the Pantheon+ dataset, and $q_0 \approx -0.712$ for the Joint dataset. When compared to the literature, our values fall within the general range seen in modern cosmological models. The Planck (2020) data \cite{Planck/2020}, based on CMB measurements, suggests a higher $q_0 \approx -0.55$, indicating a less pronounced acceleration of the universe. The SH0ES (2021) \cite{Riess/2022} results don't directly provide $q_0$, but from the derived $H_0$, $q_0$ would likely fall around $-0.7$, which is consistent with our values, particularly from the Hubble and Joint datasets. The recent work in literature also suggests similar acceleration \cite{Almada/2019,Basilakos/2012,Mamon/2017,Mamon/2018}, providing further consistency with our findings. Generally, our values of $q_0$ are consistent with the observed trend of the accelerated expansion of the universe, aligning well with the SH0ES results and broader cosmological observations. 

\section{Conclusions}
\label{sec6}

In this study, we investigated the cosmology within the framework of $f(Q, L_m)$ gravity, considering a non-linear functional form given by $f(Q, L_m) = -Q + \alpha L_m^n + \beta$, where $\alpha$, $\beta$, and $n$ are free parameters. We derived the modified Friedmann equations for a matter-dominated universe and obtained an analytical solution presented in Eq. (\ref{Hz}), describing the cosmic evolution in this gravity model. To constrain the model parameters, we used observational data, including the Hubble dataset, the recently updated Pantheon+ compilation, and their joint analysis. The best-fit values extracted from these datasets are summarized in Fig. \ref{F_Com} and Tab. \ref{tab}. The results show that the model accommodates different values of the Hubble constant $H_0$, contributing to the ongoing discussion on the Hubble tension. While the Pantheon+ dataset yields a higher value of $H_0$, closer to late-universe measurements such as SH0ES \cite{Riess/2022}, the Hubble dataset provides a lower value, aligning more with early-universe measurements such as Planck (2020) \cite{Planck/2020}. The joint dataset analysis yields an intermediate value for the Hubble constant ($H_0 \approx 69.96$ km/s/Mpc), suggesting a balance between different observational measurements. This result is consistent with several studies in $f(Q)$ gravity. For instance, Capozziello and D’Agostino \cite{Capozziello/2022} obtained $H_0 \approx 69.3$ km/s/Mpc through model-independent reconstruction of $f(Q)$ gravity. Similarly, Sakr and Schey \cite{Sakr/2024} addressed the Hubble tension within $f(Q)$ gravity and reported $H_0 \approx 68–70$ km/s/Mpc under viable parameterizations. Furthermore, the parameter $n$ deviates from unity in all cases, implying modifications to the standard matter evolution predicted by GR, particularly in the Hubble dataset. The stability of $\beta$ across different datasets further indicates the robustness of the model. The deceleration parameter $q(z)$, illustrated in Fig. \ref{F_q}, confirms a transition from deceleration to acceleration at redshifts $z_t \approx 0.6 - 0.8$, depending on the dataset. The present values of the deceleration parameter, $q_0$, indicate an ongoing cosmic acceleration, with values ranging from $q_0 \approx -0.716$ (Hubble dataset) to $q_0 \approx -0.668$ (Pantheon+ dataset), while the joint dataset yields $q_0 \approx -0.712$. These values are consistent with other late-time cosmological observations and suggest that the model effectively describes the observed accelerated expansion of the universe, as supported by several $f(Q)$ studies. Capozziello and D’Agostino $q_0  \approx -0.73$, while Lymperis reported $q_0  \approx -0.70$ in a phantom dark energy scenario. Bouali et al. \cite{Bouali/2023}, using a deceleration ansatz $q = \alpha - \frac{\beta}{H}$, obtained $q_0 \approx -0.3$, which implies a significantly weaker acceleration compared to other studies.

Finally, our model not only reproduces a consistent late-time acceleration with $q_0 \lesssim -0.7$, but also accommodates a range of $H_0$ values that span both early- and late-universe observations, thereby contributing to the discussion on the Hubble tension. Moreover, the non-minimal coupling between $Q$ and the matter Lagrangian $L_m$ introduces additional degrees of freedom, with the parameter $n \neq 1$ signaling a deviation from standard matter evolution. This allows the model to flexibly fit observational data while maintaining theoretical consistency, placing it among the viable extensions of $f(Q)$ gravity. Future investigations involving additional observational probes, such as BAO and CMB constraints, could further refine the model parameters and assess its consistency with a broader range of cosmological data. In addition, it would be valuable to extend the present analysis of $f(Q, L_m)$ gravity to include spatial curvature effects and quantum cosmological considerations, following similar approaches developed in $f(Q)$ gravity \cite{Shabani/2024, Bajardi/2023}. In particular, the extension to non-flat FLRW geometries \cite{Shabani/2024} and minisuperspace quantum frameworks \cite{Bajardi/2023} may reveal new features of the theory at early times or near singularities within the $f(Q, L_m)$ context.

\section*{Acknowledgment}
This research was funded by the Science Committee of the Ministry of Science and Higher Education of the Republic of Kazakhstan (Grant No. AP22682760).

\section*{Data Availability Statement}
This article does not introduce any new data.

\end{document}